\documentclass[12pt,a4paper]{article}
\usepackage{amsmath}
\usepackage{amssymb}
\usepackage{graphicx}
\usepackage{setspace}
\usepackage{epsfig}
\usepackage{algorithm}
\usepackage{algorithmic}
\usepackage{supertabular}
\usepackage{pstricks}
\usepackage{algorithm}
\usepackage{algorithmic}
\usepackage{supertabular}
\newcommand{\no}{\noindent}
\begin{document}
\begin{center}
{\large\bf  Idempotents, \mbox{Mattson-Solomon} Polynomials and Binary LDPC codes}\\[5mm]
R. Horan, C. Tjhai, M. Tomlinson, M. Ambroze and M. Ahmed\\
{\em Fixed and Mobile Communications Research}\\
{\em University of Plymouth, PL4 8AA, United Kingdom}\\
{\small\bf \today}
\end{center}
\begin{abstract}
\no We show how to construct an algorithm to search for binary
idempotents which may be used to construct binary LDPC codes. The
algorithm, which allows control of the key properties of
sparseness, code rate and minimum distance, is constructed in the
Mattson-Solomon domain. Some of the new codes, found by using this
technique, are displayed.
\end{abstract}
\vspace{1cm}\section{Introduction and Background} \label{intro}
\no The use of idempotents in the construction of cyclic error
correcting codes is well established and the resulting literature
is extensive (for example, see
\cite{McW},\,\cite{Rmn},\,\cite{vanL1}). The basic building blocks
for this theory are the primitive idempotents. Any cyclic code may
be described by a unique idempotent and this idempotent is a sum
of  primitive idempotents. For binary cyclic codes, efficient
algorithms exist for the calculation of these primitive
idempotents.

Another way of constructing idempotents in the binary case is by
using cyclotomic cosets and it was this property which was
exploited by  Shibuya and  Sakaniwa in \cite{Shi}.
 Their goal was  to use idempotents to construct parity check matrices for LDPC
 codes which have
 no cycles of length 4 in their factor graphs. At the heart of their technique is a
 lemma which is a variation of a result used by Weldon
 \cite{We}, for the construction of difference set cyclic codes.
 Using this lemma and a subsequent theorem,  they were able to
 simplify the problem of determining which  of the idempotents
 that are constructed, using a single cyclotomic coset, do not have cycles of length 4. They then
 extended this theory to  more general idempotents.

This approach to the construction of LDPC codes has the great
advantage of simplicity, the parity check matrices depend only
upon the correct choice of cyclotomic cosets and these are very
easily calculated. However, we believe that this advantage is
offset by some fundamental weaknesses.

Whilst the absence of 4 cycles is a desirable objective in the
construction of LDPC codes it is not mandatory \cite{Tian}, since
there are some good codes which do not have this property. An
example of such a code is included in this paper. The code rate is
also an important property of codes but, as Shibuya and Sakaniwa
admit in their conclusion, the codes which they construct in this
way are \lq\lq \textit{expected} to have a large minimum distance
\emph{at the expense of rate}\rq\rq (our italics). The minimum
distance of a code is a crucial property but there is no
indication in \cite{Shi} of how either a single cyclotomic coset,
or combinations of more than one cyclotomic cosets, should be
chosen to guarantee that the code constructed has a large minimum
distance.

 In order to address the question of how to choose
idempotents which will produce good LDPC codes we propose an
entirely different route. As in \cite{Shi}, we shall deal
exclusively with binary cyclic codes.  Making effective use of the
Mattson-Solomon polynomial, we produce an algorithm which not only
allows us to choose, in a systematic way, idempotents with low
weight, and therefore a correspondingly sparse parity check
matrix, but also with the desirable features that the
corresponding codes have a high code rate  and a large minimum
distance.

This paper is organised  as follows. In \mbox{section\
\ref{theory}} we shall review the necessary theory
 and explain how
it will be used to provide an algorithm for the determination of
idempotents which may be used to construct good codes. In
\mbox{section\ \ref{imp}} the design and implementation of this
algorithm is given and then, in \mbox{section\ \ref{exs}}, some of
the results are displayed. Finally, in \mbox{section\ \ref{con}},
we draw our conclusions on this approach.
%
%
%%%%%%%%%%%%%%%%%%%%%%%%%%%%%%%%%%%%%%%%%%%%%%%%%%%%%%%%%
%
%
\section{Binary Idempotents}\label{theory}
Let $F=GF(2)$, $n$ be a positive integer and $\mathcal{F}$ be the
splitting field for $x^n-1$ over $F$. Let $\alpha\in\mathcal{F}$
be a primitive $n$th root of unity and let $T(x)$ be the
polynomials in $\mathcal{F}[x]$ of degree $\leq n-1$. If $a(x)\in
T(x)$ then the map $\Phi\, :T\rightarrow T $ is defined by
\begin{equation}\label{MS}
[\Phi(a)](z)=\sum_{j=1}^na(\alpha^j)z^{n-j}
\end{equation}and $\Phi(a)$ is the
Mattson-Solomon polynomial of $a$ (see \cite{McW}). (We use $x$
and $z$ for the polynomial variables to distinguish between the
polynomials in the domain and codomain of $\Phi$.) If $\circ$ is
multiplication of polynomials$\mod (x^n-1)$ and $\ast$ is defined
on $T(z)$ by the rule $(\sum a_iz^i)\ast(\sum b_iz^i)=\sum
a_ib_iz^i$ then it is well known \cite{McW},\cite{vanL1}, that
$$\Phi\, :(T,+,\circ )\rightarrow (T,+,\ast )
$$ is an isomorphism of rings, in particular it is an isomorphism
of the additive groups.

If $S(x)$ is the subset of $T(x)$ consisting of polynomials with
coefficients
 in $GF(2)$ (binary polynomials) and  $E(x)$ is the subset of $T(x)$ consisting of idempotents,
both of these subsets are  \textit{additive subgroups} of $T(x)$.
It is easy to show (see \cite{McW}) that
\begin{eqnarray}\label{isos}
  \Phi\,:(S(x),+) &\rightarrow & (E(z),+) \\
  \Phi\,:(E(x),+) &\rightarrow & (S(z),+)
\end{eqnarray}are both isomorphisms and
%$$\Phi\,:(S(x),+)\rightarrow (E(z),+)$$and$$\Phi\,:(E(x),+)\rightarrow (S(z),+)$$are both isomorphisms.
from this it is obvious that
\begin{equation}
\label{main_iso}\Phi\,:(S(x)\cap E(x),+)\rightarrow (E(z)\cap
S(z),+)
\end{equation}is also an isomorphism.

\medskip Suppose that  $u(x)$ is a binary idempotent which is used to
construct a parity check matrix for a cyclic code. The parity
check matrix is constructed from the $n$-cyclic shifts of $u(x)$
\cite{Pl}, and so for the resulting code to be a LDPC code, $u(x)$
must have low weight.

If $h(x)=\mathrm{gcd}(x^n-1,u(x))$ and $g(x)=(x^n-1)/h(x)$, then
$g(x)$ is the generator of the cyclic code. If the generator,
$g(x)$, has degree $n-k$,  the dimension of the code is $k$ and
the larger the value of $k$, the better the code rate. Since
$g(x)$ is a divisor of $x^n-1$, all of the zeros of $g(x)$ are
$n$th roots of unity, and there are $n-k$ of these. Further,
gcd$(g(x),h(x))$=1 and $x^n-1=h(x)g(x)$, so that the number of
distinct $n$th roots of unity which are also roots of $u(x)$ is
$k$. The dimension of the code is therefore the number of $n$th
roots of unity which are also roots of $u(x)$.

The BCH bound of the code is determined by the number of
consecutive powers of $\alpha$, taken cyclically (mod\,$ n$),
which are also roots of $g(x)$. For the reasons outlined in the
previous paragraph, this is precisely the same as the number of
consecutive powers of $\alpha$, taken cyclically (mod\,$ n$),
which are \textit{not} roots of $u(x)$.

The important features of the code are therefore determined by:

\medskip
\begin{tabular}{cl}
 (a) & the weight of the idempotent $u(x)$, \\
  (b) & the number of $n$th roots of unity which are roots of $u(x)$, \\
   (c) & the number of consecutive
powers of $\alpha$ which are \textit{not} roots of $u(x)$.
\end{tabular}

\medskip Take $u(x)\in S(x)\cap E(x)$ and let $\Phi(u)=\theta$ be
its MS polynomial. The inverse mapping

\begin{equation}
\label{inv_iso}\Phi^{-1}\,:(S(z)\cap E(z),+)\rightarrow (E(z)\cap
S(x),+)
\end{equation}is defined as follows:
If $A(z)=[\Phi(a)](z)$ is the Mattson-Solomon polynomial of the
polynomial $a(x)=a_0+a_1x+\dots+a_nx^{n-1}$ then, for $i=0,\ldots
,n-1,$
\begin{equation}\label{inverse}
a_i=\frac{1}{n}A(\alpha^i)
\end{equation}
(see \cite{McW}). Let $h(z)=\mathrm{gcd}(\theta(z),z^n-1)$ and let
$f(z)=(z^n-1)/h(z)$.
% $f(z)\in GF(2)[z]$ be the divisor of $x^n-1$ which is the
%check polynomial for the cyclic code generated by $\theta(z)$.
The three key properties relating to the idempotent $u(x)$, listed
above, are easily gleaned from its Mattson-Solomon polynomial
$\theta (z)$, and $f(z)$, as follows:

%
%%%
%%%%%%%%%%%%%%%%%%%%%%%%%%%%%%%%%%%%%%%%%%%%%%%%%%%%%%%%%%%%%
%
%
\subsection{The weight of $\mathbf{ u(x)}$}\label{wt-of-u}

The weight of $u(x)$ is the number of  $n$th roots of unity which
are zeros of $f(z)$. To see this note that  $f(\alpha^i)=0$ if and
only if $\theta(\alpha^i)=1$, since idempotents take only the
values $0$ and $1$ in $ \mathcal{F}$. Now $u=\Phi^{-1}\theta$ and
the coefficients of $u(x)=u_0+u_1x+\ldots+u_{n-1}x^{n-1}$ are
given by
\begin{equation}\label{coeffs}
u_i=\theta(\alpha^i) \mod 2\quad \mathrm{for}\ i=0,\ldots
n-1\,
 \end{equation} (\textit{cf } equation (\ref{inverse})). Thus
  $ u_i=1$ precisely
when $f(\alpha^i)=0$, giving the weight of $u(x)$ as the degree of
the polynomial $f(z)$.

%\medskip\no\textbf{(b) The zeros of} $\mathbf{ u(x)}$.
\subsection{The zeros of $\mathbf{ u(x)}$.}\label{zeros-of-u}
From the definition of the MS polynomial,
 (\ref{MS}),
\begin{equation}\label{theta}
\theta(z)=\sum_{j=1}^nu(\alpha^j)z^{n-j}
\end{equation}
and
   the number of zeros of $u(x)$ which are roots of unity is clearly $n-\mathrm{wt}(\theta(z))$.

%\medskip\no\textbf{(c) The BCH bound of the code.}
\subsection{The BCH bound of the code.}\label{bch-bound}
The BCH bound of the code is the largest number of consecutive
powers of $\alpha$ which are \textit{not} roots of $u(x)$, i.e.
the number of consecutive $i$, taken (mod\,$ n$), such that
$u(\alpha^i)=1$. From (\ref{theta}), this is the largest number of
consecutive non-zero coefficients in $\theta$, taken cyclically
(mod $n$).

\bigskip\no Using this information, a systematic search for
idempotents can now be made in increasing order of weight, with
accompanying knowledge of the number of roots which are $n$th
roots of unity and the corresponding BCH bound. This algorithm is
constructed in the Mattson-Solomon domain.

Let the decomposition of $z^n-1$ into
 irreducible (over $F=GF(2)$) polynomials be $z^n-1=f_1(z)f_2(z)\ldots
 f_t(z)$.
 For $i=1,\ldots,t,$ let $k_i(z)=(z^n-1)/f_i(z)$ and let $\theta_i(z)$ be the associated
 primitive idempotent (see \cite{McW} or \cite{vanL1}).
These are displayed below in an array, together with other
idempotents:

\bigskip
\begin{equation}\label{array1}
\left.
\begin{array}{cccc}%\hline
  u_1(x) & \null \qquad\null & \theta_1(z) & f_1(z) \\
  u_2(x) & & \theta_2(z) & f_2(z) \\
  \vdots &  & \vdots & \vdots \\
  u_t(x) &  & \theta_t(z) & f_t(z)\\
  %\hline
\end{array}
\right\}
\end{equation}

\bigskip
\no Here  $u_1(x),u_2(x),\ldots,u_t(x)$ are the idempotents whose
Mattson-Solomon polynomials are $\theta_1(z),\theta_2(z),\ldots
,\theta_t(z)$, respectively. Let $I\subseteq \{1,2,\ldots,t\}$ and
let $u,\theta$ and $f$ be defined as $u=\sum_{i\in I}u_i$,
$\theta=\sum_{i\in I}\theta_i$ and $f(z)=\prod_{i\in I}f_i(z)$.
From the properties of primitive idempotents, if
$h(z)=\mathrm{gcd}(\theta(z),z^n-1)$ then it follows that
gcd$(f(z),h(z))=1$ and $z^n-1=f(z)h(z).$ The idempotent $u$ will
now have the following properties.
\begin{eqnarray}
   \mathrm{wt}(u) &=& \sum_{i\in I}\mathrm{deg}(f_i)\label{eqn:wt-of-u}\,,\\
\mathrm{number\ of\ zeros\ of\ u}&=& n-\mathrm{wt}(\theta)\label{eqn:zeros-of-u}\,.
\end{eqnarray}
The BCH bound is determined from $\theta(z)$ as explained in
\ref{bch-bound}.

Since methods for finding the $\theta_i$ and $f_i$ are  well
documented (see e.g.
 \cite{vanL2})
 a search algorithm can be built
 around this observation to find a suitable weight idempotent with a known  number of zeros and a known BCH bound.
 The rows of the array (\ref{array1}), are ordered  by the degree of the  polynomials, i.e.
 $\mathrm{deg}(f_i)\leq\mathrm{deg}(f_{i+1})$ for all $i$, and a search can be made in increasing order of weight.
%The \textit{number} of zeros  is easy to find at each stage.
When a successful outcome has been obtained, \textit{only at this
stage}
%is it necessary to take the \textit{inverse Fourier transform}
is the \textit{inverse Fourier transform} (MS$^{-1}$)  evaluated
to find the corresponding idempotent. All of the information which
is required will already be known.

\section{Design and Implementation}\label{imp}
If $t$ denotes the number of cyclotomic cosets modulo $n$, the
complexity of an exhaustive search algorithm is
$\mathcal{O}(2^t)$. We reduce this search complexity by targeting
the search on the three key parameters:

%\begin{enumerate}
\subsection{Sparseness of the parity-check
matrix}\label{sparseness}
 In \cite{We}, Weldon introduced
difference-set cyclic codes. These codes have the desirable
property that the parity check equations are orthogonal on all
bits and  have no cycles of length 4 in their factor graphs. A
necessary condition for this is that if $v(x)$ is the polynomial
which generates the parity check matrix then the weight of $v(x)$
must satisfy the inequality
\begin{equation}\label{eqn:diff-set}
\mathrm{wt}(v(x))(\mathrm{wt}(v(x))-1)\leq n\,,
\end{equation}where $n$ is the code length. Since the weights of the idempotents
 $u(x)$ are related  to the
degrees of the $f_i$ by (\ref{eqn:wt-of-u}), a reasonable bound is

\begin{align}
\sum_{i\subseteq I}\text{deg}(f_i) \leq
\sqrt{n}\,.\label{eqn:wt-bound}
\end{align}In practice we have gone a little beyond this limit
and this has enabled us to find some good codes which do have
cycles of length 4 in their factor graph.

\subsection{Code-rate}\label{code-rate}
The code-rate is directly proportional to the number of roots of
$u(x)$. If we let $R_{min}$ represent our minimum desired
code-rate then, following equation~(\ref{eqn:zeros-of-u}), we can
refine our search bound to
\begin{align}
 \text{wt}(\theta) \le (1-R_{min})n\,.\label{eqn:theta-bound}
\end{align}

\subsection{Minimum distance}\label{min-dist}
Let $d$ be the lowest desired minimum distance and let
$r_{\theta}$ be the largest number of consecutive non-zero
coefficients, taken cyclically $\mod  n$, of $\theta$. Then,
following the discussion of \ref{bch-bound}, we restrict our
search algorithm to those $\theta$ for which

\begin{align}
r_{\theta} > d\label{eqn:dmin-bound}
\end{align}

\noindent We develop an efficient, but exhaustive recursive
tree-search based on the above bounds. The developed search
algorithm, Algorithm~\ref{alg:search}, is initialised by setting
$\mathbf{V}$ and $index$ to $\emptyset$ and $-1$ respectively.
% \newpage
\renewcommand{\algorithmicrequire}{\textbf{Input:}}
\renewcommand{\algorithmicensure}{\textbf{Output:}}
\begin{algorithm}
\caption{CodeSearch($\mathbf{V}$, $index$, $\mathbf{F}(x)$, $\mathbf{Q}(z)$)}\label{alg:search}
\begin{algorithmic}[1]
\REQUIRE $R_{min} \Leftarrow$ minimum code-rate of interest\\
$d \Leftarrow$ lowest expected minimum distance\\
$\delta \Leftarrow$ small positive integer\\
$\mathbf{F}(x) \Leftarrow \{f_i(x)\}$ $\forall i \in I$ sorted in ascending order of the
degree\\$\mathbf{Q}(z) \Leftarrow \{\theta_i(z)\}$ $\forall i \in I$
\ENSURE $\mathbf{CodesList}$ contains set of codes
\STATE $\mathbf{T} \Leftarrow \mathbf{V}$
\FOR {\big($i$=$index$+1; $i \le \text{Size}\left(I\right)$; $i$++\big)}
\STATE $\mathbf{T}_\text{prev} \Leftarrow \mathbf{T}$
\IF {\big($\sum_{\forall j \in\mathbf{T}}\text{deg}(f_j(x))+\text{deg}(f_i(x)) \le \sqrt{n}+\delta$\big)}
\STATE Append $i$ to $\mathbf{T}$
%\STATE $\text{W} \Leftarrow \sum_{\forall j \in \mathbf{T}}\text{deg}\left(f_j(x)\right)$
\STATE $\theta(z) \Leftarrow \sum_{\forall j \in \mathbf{T}} \theta_j(z)$
\IF {\big($\text{wt}(\theta(z)) \le (1-R_{min})n$ \textbf{and} $r_{\theta} > d\big)$}
\STATE $u(x) \Leftarrow \text{MS}^{-1}\left(\theta(z)\right)$
\IF {$u(x)$ is non-degenerate}
%\footnote{If there is a common factor between $n$
%and all exponents of $u(x)$ apart from unity, the cyclic code defined by $u(x)$
%is a degenerate code}
\STATE $\mathcal{C} \Leftarrow$ a cyclic code defined by $u(x)$
\IF {\big($\mathcal{C} \notin \mathbf{CodeList}$\big)}
\STATE Add $\mathbf{C}$ to $\mathbf{CodeList}$
\ENDIF
\ENDIF
\ENDIF
\STATE CodeSearch($\mathbf{T}$, $index$, $\mathbf{F}(x)$, $\mathbf{Q}(z)$)
\ENDIF
\STATE $\mathbf{T} \Leftarrow \mathbf{T}_\text{prev}$
\ENDFOR
\end{algorithmic}
\end{algorithm}
%\newpage
\section{Code Example and Performance}\label{exs}
Since the algorithm is an exhaustive search, the code construction
method presented in this paper is able to produce, in addition to
 new codes, many well-know cyclic codes, for example the
Difference-Set Cyclic codes and the Euclidean and Projective
Geometry codes. Some of the new codes which we have found using
this technique are presented in Table~\ref{tbl:code-examples}. All
codes in Table~\ref{tbl:code-examples}, except those labelled with
$^*$, have orthogonal parity-check equations.

\begin{center}
\bottomcaption{\label{tbl:code-examples}Examples of the constructed codes}
\tablefirsthead{\hline\hline
    \multicolumn{1}{|c|}{($n$, $k$)} &
    \multicolumn{1}{|c|}{$u(x)$} &
    \multicolumn{1}{|c|}{$d_{min}$}\\
\hline\hline}
\tablehead{\hline
    \multicolumn{3}{|l|}{\small \tablename\ \thetable{} $-$ \sl continued from previous page}\\\hline
    \multicolumn{1}{|c|}{($n$, $k$)} &
    \multicolumn{1}{|c|}{$u(x)$} &
    \multicolumn{1}{|c|}{$d_{min}$}\\
\hline}
\tabletail{\hline
    \multicolumn{3}{|r|}{\small \sl continue on next page}\\
\hline}
\tablelasttail{\hline\hline}
\begin{supertabular}{|l@{\hspace{0.1in}}|p{3.5in}|r@{\hspace{0.2in}}|}
(51,26)$^*$ & $1+x^3+x^6+x^{12}+x^{17}+x^{24}+x^{27}+x^{34}+x^{39}+x^{45}+x^{48}$ & $10$\\\hline
(63,44)$^*$ & $1+x^7+x^9+x^{14}+x^{18}+x^{27}+x^{28}+x^{35}+x^{36}+x^{45}+x^{49}+x^{54}+x^{56}$ & $8$\\\hline
(93,47) &  $1+x^2+x^8+x^{31}+x^{32}+x^{35}+x^{47}$ & $8$\\\hline
(105,53) & $ 1+x^4+x^{30}+x^{32}+x^{45}+x^{46}+x^{53}$ & $8$\\\hline
(117,72)$^*$ & $1+x+x^2+x^4+x^8+x^{11}+x^{16}+x^{22}+x^{32}+x^{44}+x^{59}+x^{64}+x^{88}$ & $12$\\\hline
(127,84)$^*$ & $1+x+x^2+x^4+x^8+x^{16}+x^{32}+x^{55}+x^{59}+x^{64}+x^{91}+x^{93}+x^{109}+x^{110}+x^{118}$ &
$10$\\\hline
(219,101) & $1+x^2+x^8+x^{32}+x^{73}+x^{74}+x^{77}+x^{89}+x^{110}+x^{128}+x^{137}$ & $12$\\\hline
(255,135) & $1+x^4+x^{13}+x^{21}+x^{39}+x^{54}+x^{55}+x^{91}+x^{121}+x^{123}+x^{148}+x^{195}$ & $13$\\\hline
(255,175) & $1+x+x^3+x^7+x^{15}+x^{26}+x^{31}+x^{53}+x^{63}+x^{98}+x^{107}+x^{127}+x^{140}+x^{176}+x^{197}+
x^{215}$ & $17$\\\hline
(341,205) & $1+x^{29}+x^{87}+x^{92}+x^{94}+x^{114}+x^{122}+x^{156}+x^{202}+x^{203}+x^{213}+x^{217}+x^{234}+
x^{257}+x^{273}$ & $16$\\\hline
(511,199) & $1+x+x^3+x^7+x^{15}+x^{31}+x^{63}+x^{82}+x^{100}+x^{127}+x^{152}+x^{165}+x^{201}+x^{255}+x^{296}+
x^{305}+x^{331}+x^{403}$ & $19$\\\hline
(511,259) & $1+x^{31}+x^{42}+x^{93}+x^{115}+x^{217}+x^{240}+x^{261}+x^{360}+x^{420}+x^{450}+x^{465}$ &
$13$\\\hline
(819,435) & $1+x+x^3+x^7+x^{15}+x^{31}+x^{63}+x^{127}+x^{204}+x^{255}+x^{409}+x^{511}$ & $13$\\\hline
(819,447) & $1+x+x^3+x^7+x^{15}+x^{31}+x^{63}+x^{127}+x^{204}+x^{255}+x^{350}+x^{409}+x^{511}+x^{584}+x^{701}$
& $16$\\\hline\hline
\end{supertabular}
\end{center}

\begin{figure}[hbt]
\centering\includegraphics[angle=270,width=4in]{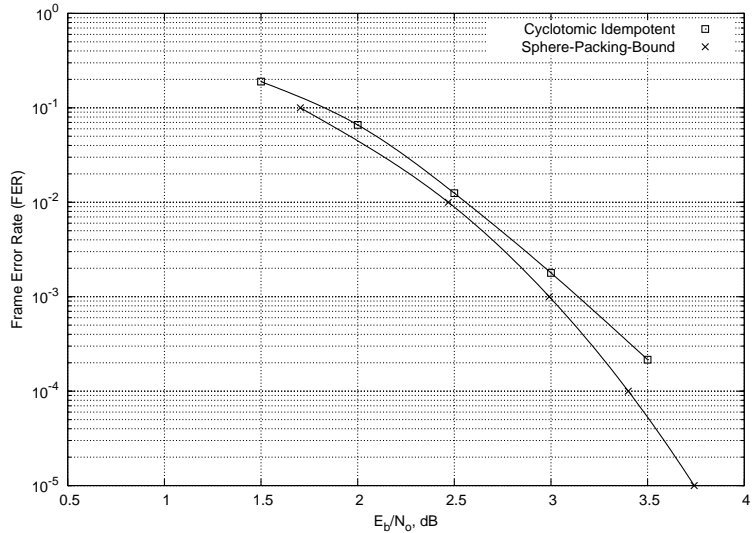}
\caption{\label{fig:fer-gf2-127-84}Frame error performance of the
$(127,84)$ cyclic code}
\end{figure}

\no Throughout the paper, it is assume that the codewords are
transmitted across a noisy communication channel with BPSK
modulation and at the receiving end is the modified
Belief-Propagation decoder which approximates the
Maximum-Likelihood decoder~\cite{Patent}.

Figure~\ref{fig:fer-gf2-127-84} shows the frame-error-rate (FER)
performance of the $(127,84)$ cyclic code, which is a code which
does have cycles of length 4.  Neverthless, the performance of
this code is outstanding and, at $10^{-3}$ FER, it is within
$0.2$dB  of the sphere-packing-bound constraint for binary
transmission.

\begin{figure}[hbt]
\centering\includegraphics[angle=270,width=4in]{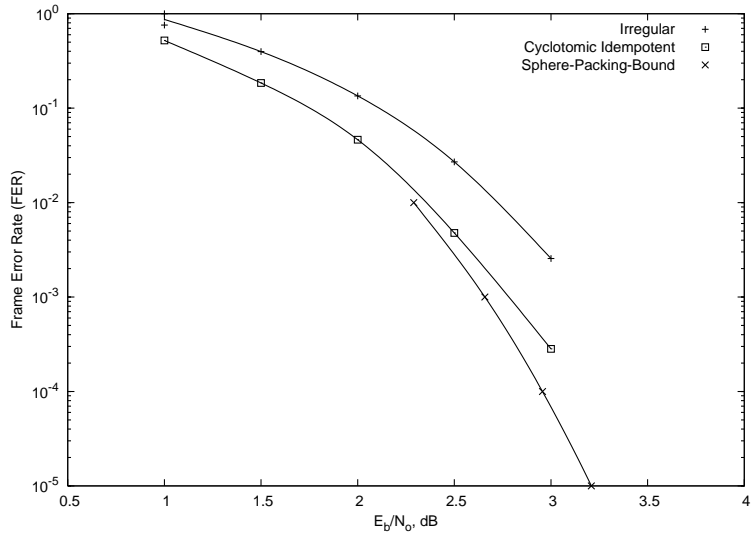}
\caption{\label{fig:fer-gf2-255-175}Frame error performance of the
$(255,175)$ codes}
\end{figure}

Figure~\ref{fig:fer-gf2-255-175} shows the performance of two
$(255,175)$ codes, one constructed using our method and one an
irregular computer generated code. We can see that our code, which
achieves a coding gain of around $0.4$dB compared to the
equivalent irregular code, performs approximately withinr $0.15$dB
of the sphere-packing-bound constraint for binary transmission.

\begin{figure}[hbt]
\centering\includegraphics[angle=270,width=4in]{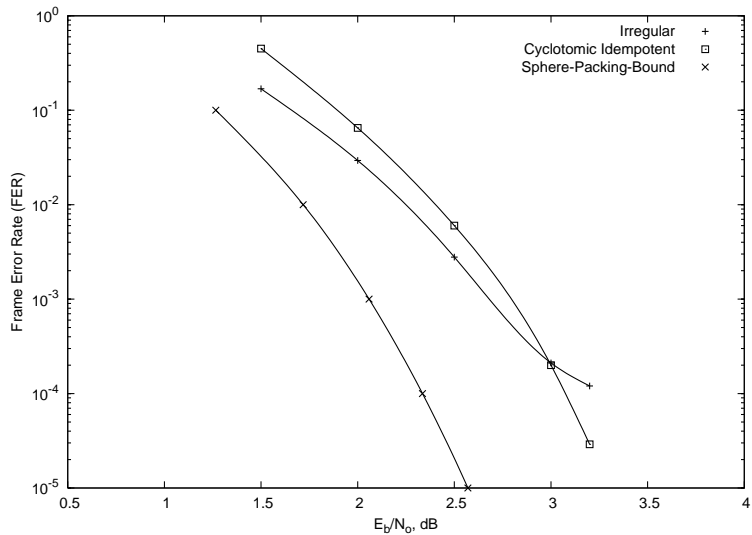}
\caption{\label{fig:fer-gf2-341-205}Frame error performance of the
$(341,205)$ codes}
\end{figure}

Our construction method can produce LDPC codes with high minimum
distance and therefore they do not suffer from error-floor.
Figure~\ref{fig:fer-gf2-341-205} demonstrates the performance of
our $(341,205)$ code, which is inferior to the equivalent
irregular code in the low signal-to-ratio region, but the
irregular code exhibits early error-floor due to its low minimum
distance.

\section{Conclusions}\label{con}
A method of constructing binary cyclic codes from the finite-field
transform (Mattson-Solomon) domain is able to produce a large
number of codes which have high minimum-distance and code-rate.
These codes have sparse parity-check matrix and thus are
applicable as LDPC codes. Due to their cyclic property these LDPC
codes have $n$ parity-check equations instead of $n-k$ equations
as in the case of random LDPC codes. With these extra parity-check
equations to iterate with, the performance of the iterative
decoder is improved.

In designing cyclic LDPC codes of length $n$, our new method
allows one to increase the $d_{min}$ of the code by combining
additional irreducible factors of $1+z^n$ which, in turn, reduces
the sparseness of the parity-check matrix. The ability to control
the sparseness of the parity-check matrix is a trade-off against
the minimum distance of the code.

Simulation results have shown that the our cyclic codes have
outstanding performance which are superior to the equivalent
irregular LDPC codes. The high $d_{min}$ of these codes ensures
the absence of an early error-floor in their performance.

\end{document}